\documentclass[aps,prl,reprint,amsmath,amssymb,superscriptaddress,longbibliography]{revtex4-2}
\usepackage{graphicx}
\usepackage{booktabs}
\usepackage{bm}
\usepackage[colorlinks=true,citecolor=blue,linkcolor=blue,urlcolor=blue]{hyperref}
\newcommand{\avg}[1]{\left\langle #1\right\rangle}
\newcommand{\dd}{\mathrm{d}}

\newcommand{\Ufour}{\Upsilon_{4}}
\begin{document}
\title{Altermagnetic Control of Charge-\texorpdfstring{\(4e\)}{4e} Superconductivity}
\author{Yan-Guang Yue}
\affiliation{Department of Physics and State Key Laboratory of Surface Physics, Fudan University, Shanghai 200433, China}
\author{Jie Lou}
\affiliation{Department of Physics and State Key Laboratory of Surface Physics, Fudan University, Shanghai 200433, China}
\author{Yan Chen}
\email{yanchen99@fudan.edu.cn}
\affiliation{Department of Physics and State Key Laboratory of Surface Physics, Fudan University, Shanghai 200433, China}
\affiliation{Shanghai Branch, Hefei National Laboratory, Shanghai 201315, China}
\begin{abstract}
We show that a two-dimensional classical array of conventional superconducting islands coupled by altermagnetic Josephson junctions supports an independent charge-\(4e\) phase.
Tuning the altermagnetic splitting suppresses the leading single-pair tunneling harmonic while retaining strong double-pair coupling, producing a finite charge-\(4e\) region between staggered and uniform charge-\(2e\) states.
Using the complete temperature-dependent microscopic Josephson spectrum, we find short-range ordinary pair correlations alongside algebraic coherence of pairs of Cooper pairs.
Berezinskii--Kosterlitz--Thouless scaling and a half-flux response further identify the charge-\(4e\) phase.
Our results establish altermagnetism as a microscopic means to select between ordinary and composite superconducting coherence.
\end{abstract}
\maketitle

\textit{Introduction.---}
In an independent charge-\(4e\) superconductor, pairs of Cooper pairs retain coherence after ordinary Cooper-pair coherence is lost.
Proposed routes include melting pair-density-wave order \cite{four_Agterberg2008,four_Berg2009,four_BergTheory2009}, fluctuations of multicomponent superconductivity \cite{four_Herland2010,four_Fernandes2021,four_Jian2021,four_Volovik2024}, and interacting electronic models \cite{four_Jiang2017,four_Gnezdilov2022,four_Gao2026primary}.
Josephson networks provide another setting: frustrated chains and protected circuits support paired transport \cite{four_DoucotVidal2002,four_IoffeFeigelman2002,four_ProtopopovFeigelman2004,four_Gladchenko2009,four_Bell2014}, and a recent dice-array proposal realizes collective quartet correlations using imposed flux \cite{four_Weerda2026}.
Magnetic control also has a direct precedent.
Near a ferromagnetic junction's \(0\)--\(\pi\) crossover \cite{meth_Buzdin2005}, a two-harmonic array model supports an intermediate nematic \(4e\) phase and half-flux periodicity \cite{meth_Korshunov2010}.
The microscopic harmonic spectrum is therefore decisive: suppressing its first harmonic need not suppress the ordinary coherence supported by all remaining odd terms.

Altermagnets combine compensated magnetic order with momentum-dependent spin splitting \cite{am_smejkal2022symmetry,am_smejkal2022landscape}, observed in candidate materials \cite{am_krempasky2024,am_reimers2024}.
Superconducting hybrids \cite{am_fukaya2025review} exhibit phase-shifted Andreev states and \(0\)--\(\pi\) transitions \cite{am_ouassou2023,am_beenakker2023}.
They also support finite-momentum proximity pairing \cite{am_zhang2024pairing} and tunable higher harmonics \cite{am_lu2024phi,am_fukaya2025josephson,am_sun2025harmonic}.
The resulting double-pair tunneling has motivated qubit proposals \cite{am_guo2026qubit,am_pan2026qubit}.
An independent charge-\(4e\) phase requires algebraic correlations of \(e^{2i\theta}\) and short-range correlations of \(e^{i\theta}\).
This distinction also separates the present question from second-harmonic transport in semiconductor junction devices \cite{four_Ciaccia2024,four_Leblanc2024}.

\begin{figure}[tb]
\includegraphics[width=\columnwidth]{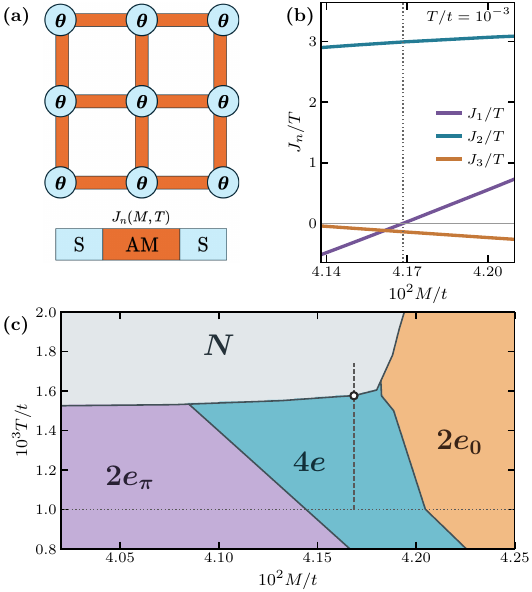}
\caption{\label{fig:phase}
Altermagnetic control of the coherent charge sector.
(a) Conventional superconducting islands (blue) connected through altermagnetic links (orange); one S/AM/S junction is shown below.
(b) Leading microscopic energy harmonics at \(T/t=10^{-3}\); the network retains the converged harmonic spectrum.
(c) Phase map: \(2e_\pi\), \(4e\), \(2e_0\), and \(N\) denote staggered ordinary, independent composite, uniform ordinary, and normal regimes.
Here \(N\) denotes the phase-incoherent regime of the array; the island pairing amplitudes remain fixed.
Interfaces interpolate between supported finite-size assignments.
The horizontal and vertical lines mark the representative temperature and fixed-\(M_*\) cuts; the white point marks the large-size BKT estimate.
Here \(M_*\) is defined by \(J_1(M_*,10^{-3}t)=0\).}
\end{figure}

Here we establish a microscopic altermagnetic route to an independent charge-\(4e\) phase in a two-dimensional classical Josephson array.
Tuning the altermagnetic splitting stabilizes a finite \(4e\) region between staggered and uniform \(2e\) states [Fig.~\ref{fig:phase}].
The full temperature-dependent Josephson spectrum supports short-range ordinary correlations alongside algebraic composite coherence, despite residual odd harmonics.
Berezinskii--Kosterlitz--Thouless (BKT) scaling and an emergent \(h/4e\) flux response identify the charge-\(4e\) phase, which survives open junction edges and weak microscopic disorder.
Altermagnetism thus controls whether Cooper pairs or their pairs remain coherent across the array.

\textit{Microscopic altermagnetic control.---}
Altermagnetic spin splitting can suppress the first Josephson harmonic while preserving a substantial second harmonic.
We consider a square array of conventional superconducting islands linked by superconductor--altermagnet--superconductor (S/AM/S) junctions [Fig.~\ref{fig:phase}(a)].
Here S and AM denote the superconducting electrodes and altermagnetic region, respectively.
The spin-conserving altermagnetic region is described by
\begin{equation}
\begin{split}
h_{\rm AM}(\bm k)={}&[t(\cos k_x+\cos k_y)-\mu_{\rm AM}]s_0\\
&+M(\cos k_x-\cos k_y)s_z.
\end{split}
\label{eq:am}
\end{equation}
Here \(t\) sets the electronic energy scale, \(\mu_{\rm AM}\) is the chemical potential, and \(s_0,s_z\) are the identity and Pauli matrices in spin space.
The parameter \(M\) controls the compensated spin splitting.
Varying \(M\) changes spin-dependent propagation through the junction and thereby the coupling between its superconducting electrodes.

For electrode phase difference \(\delta\), integrating out the junction electrons and rigid \(s\)-wave electrodes gives \(E_J(\delta)=-\sum_{n\ge1}J_n(M,T)\cos(n\delta)+{\rm const}\).
Treating junctions as independent bonds between islands of fixed pairing amplitude gives the effective phase Hamiltonian
\begin{equation}
H_{\rm net}=-\sum_{\langle ij\rangle}\sum_{n\ge1}
J_n(M,T)\cos[n(\theta_i-\theta_j)].
\label{eq:network}
\end{equation}
Each \(\theta_i\) is a thermally fluctuating island phase, and \(L\) counts islands along each direction.
We recompute the couplings at every \((M,T)\) and use the same temperature in the array's Boltzmann weight.
Both bond directions use the same microscopic spectrum.
The lattice spacing and \(k_B\) are unity.
Microscopic parameters, convergence checks, and derivations are given in the Supplemental Material (SM) \cite{supplement}.

Increasing \(M\) changes \(J_1\) from negative to positive, switching its preferred phase difference from \(\pi\) to \(0\) [Fig.~\ref{fig:phase}(b)].
Positive \(J_2\) favors both choices and locks only the doubled phase \(2\delta\).
At \(T/t=10^{-3}\), the root \(J_1(M_*,T)=0\) occurs at \(M_*/t\simeq4.17\times10^{-2}\), where \(J_2/T=2.99\).
Residual odd harmonics retain a preference between these choices: \(J_3/T=-0.14\), and their total absolute strength is \(\sum_{n\ {\rm odd}}|J_n|/T=0.17\).
These residual couplings are retained along with the higher even harmonics.
This hierarchy of Josephson harmonics favors independent charge-\(4e\) coherence.

\textit{An independent charge-\(4e\) phase.---}
We test whether the microscopic coupling hierarchy supports independent composite coherence by comparing \(C_{2e}(\bm r)=\avg{e^{i(\theta_{\bm r}-\theta_0)}}\) and \(C_{4e}(\bm r)=\avg{e^{2i(\theta_{\bm r}-\theta_0)}}\).
An independent \(4e\) phase requires short-range ordinary correlations and algebraic composite correlations; ordinary phases retain algebraic pair correlations, either uniform or staggered.
The phase map in Fig.~\ref{fig:phase}(c) identifies a finite composite region between the two ordinary states.
At \(T/t=10^{-3}\), increasing \(M\) drives \(2e_\pi\to4e\to2e_0\), demonstrating altermagnetic control over the coherent charge sector.
Heating the central region gives a normal phase with both correlations short ranged.

\begin{figure}[tb]
\includegraphics[width=\columnwidth]{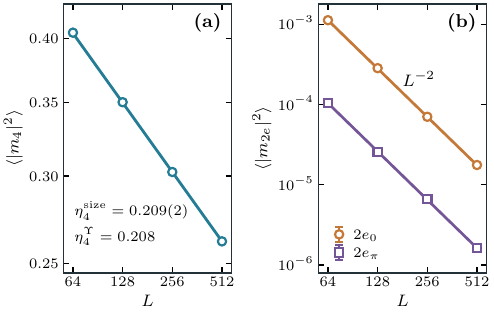}
\caption{\label{fig:independent}
Independent composite coherence at \(M=M_*\), \(T/t=1.55\times10^{-3}\).
(a) The composite moment decays algebraically for \(L=64,128,256,512\); its effective exponent from \(L=256,512\) agrees with the stiffness estimate at \(L=512\).
(b) Both uniform and staggered ordinary moments scale as \(L^{-2}\), consistent with finite ordinary structure factors.
Lines show the corresponding powers; error bars are smaller than several symbols.}
\end{figure}

Large-system scaling establishes the separation of ordinary and composite coherence.
Define \(m_4=L^{-2}\sum_i e^{2i\theta_i}\), \(m_{2,0}=L^{-2}\sum_i e^{i\theta_i}\), and \(m_{2,\pi}=L^{-2}\sum_i(-1)^{x_i+y_i}e^{i\theta_i}\).
These probe doubled-phase coherence and uniform or staggered ordinary coherence, respectively.
Their squared amplitudes are invariant under a global phase rotation.
For each channel, the structure factor obeys \(S=L^2\avg{|m|^2}=\sum_{\bm r}C(\bm r)\), with the staggered factor included for \(m_{2,\pi}\).
A finite correlation length makes this sum saturate at large \(L\), yielding \(\avg{|m|^2}\propto L^{-2}\).
For \(C(r)\propto r^{-\eta}\) with \(0<\eta<2\), the sum grows as \(S\propto L^{2-\eta}\), giving \(\avg{|m|^2}\propto L^{-\eta}\) \cite{meth_FisherBarber1972}.
At \(M=M_*\), \(T/t=1.55\times10^{-3}\), the composite moment follows algebraic scaling through \(L=512\), with \(\eta_4\simeq0.21\) [Fig.~\ref{fig:independent}(a)].
Both ordinary moments follow short-range scaling, with exponents \(2.00(2)\) and \(2.02(2)\); their structure factors saturate [Fig.~\ref{fig:independent}(b)].
In the ordinary wings, both charge sectors share the same long-wavelength phase field, giving \(\eta_4\simeq4\eta_2\).
The central region instead retains composite quasi-long-range order after both ordinary channels become short ranged.

The phase field separates into a sign and a doubled phase, \(e^{i\theta_i}=s_i e^{i\phi_i/2}\), with \(\phi_i=2\theta_i\bmod2\pi\) and \(s_i=\pm1\).
The two signs distinguish original phases differing by \(\pi\), which have identical \(e^{2i\theta_i}\).
Even harmonics depend only on \(\phi\), whereas the odd contribution becomes \(-\sum_{\langle ij\rangle}K_{ij}(\phi)s_is_j\).
At fixed \(\{\phi_i\}\), this is an Ising model with \(K_{ij}(\phi)=\sum_{n\ {\rm odd}}J_n\cos[n(\phi_i-\phi_j)/2]\); the signs and doubled phase remain coupled through these bonds.
Since each cosine has magnitude at most unity, \(|K_{ij}(\phi)|\le J_{\rm odd}=\sum_{n\ {\rm odd}}|J_n|\), independently of \(\{\phi_i\}\).
On the square lattice with four neighbors per site, a sufficient condition for exponentially decaying sign correlations is \cite{meth_Dobrushin1968}
\begin{equation}
\mathcal D=4\tanh(J_{\rm odd}/T)<1.
\label{eq:dobrushin}
\end{equation}
This condition bounds the sign correlations for every \(\{\phi_i\}\).
The ordinary correlator is the average over \(\phi\) of \(\avg{s_is_j}_{s|\phi}e^{i(\phi_i-\phi_j)/2}\).
The phase factor has unit modulus, so this average preserves exponential decay in both ordinary channels.
The composite correlator \(\avg{e^{i(\phi_i-\phi_j)}}\) contains no sign factor and can retain algebraic coherence.
At the lower representative temperature \(T/t=10^{-3}\), \(M=M_*\), \(\mathcal D=0.65\), guaranteeing exponentially decaying ordinary correlations in a central subinterval.
Together with the numerical composite correlations, this establishes coexistence of sign disorder and algebraic coherence of the doubled phase despite residual odd harmonics.

\textit{Thermal loss of composite coherence.---}
We next locate the thermal transition from the independent charge-\(4e\) phase to the normal network.
The coherent field is \(e^{i\phi}\), with \(\phi=2\theta\): its elementary \(2\pi\) vortex corresponds to a \(\pi\) winding of \(\theta\), defining a half vortex of the island phase.
We therefore test whether heating destroys composite coherence through a BKT transition driven by half-vortex unbinding.

We measure the island-phase stiffness \(\Upsilon_\theta\), the coefficient of the quadratic free-energy cost of a slowly varying \(\theta\).
Since \(\nabla\phi=2\nabla\theta\), the corresponding composite-phase stiffness is \(\Ufour=\Upsilon_\theta/4\).
Within this independent charge-\(4e\) phase, the decay exponent \(\eta_4\) of \(C_{4e}(r)\) obeys \(\eta_4=T/(2\pi\Ufour^R)\), where \(R\) denotes the long-distance renormalized stiffness.
At the transition temperature \(T_c\), half-vortex unbinding predicts \(\Ufour^R(T_c^-)/T_c=2/\pi\) on the coherent side, and hence \(\eta_4(T_c)=1/4\) \cite{meth_Berezinskii1971,meth_KosterlitzThouless1973,meth_Kosterlitz1974,meth_NelsonKosterlitz1977}.
The stiffness estimate \(\eta_4\simeq0.21\) at \(T/t=1.55\times10^{-3}\) agrees with the correlation scaling in Fig.~\ref{fig:independent}.

\begin{figure}[tb]
\includegraphics[width=\columnwidth]{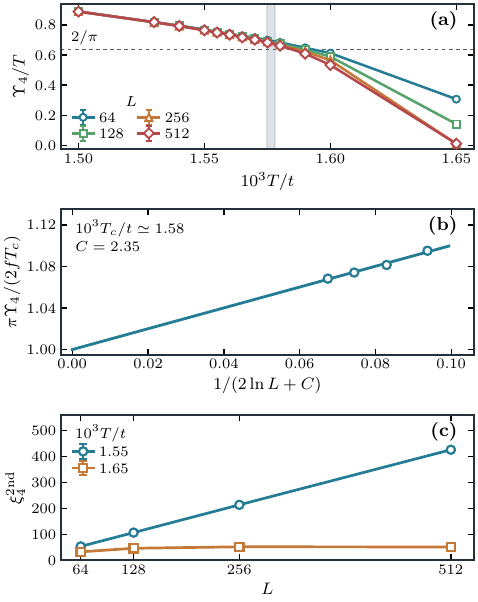}
\caption{\label{fig:bkt}
Thermal loss of composite coherence at fixed \(M=M_*\).
(a) Composite-phase stiffness \(\Ufour/T\); the horizontal line is \(2/\pi\), and the band marks the working transition interval.
(b) Leading logarithmic finite-size fit, including the square-torus factor \(f\).
The universal jump is a theoretical input to this fit.
(c) The second-moment composite correlation length \(\xi_4\) grows with \(L\) below the transition and saturates at \(T/t=1.65\times10^{-3}\).
Fit variants and winding-sector sampling limitations are reported in the SM.}
\end{figure}

We test the thermal transition at fixed \(M=M_*\) using 13 temperatures and \(L=64,128,256,512\), recomputing the full microscopic \(J_n(M_*,T)\) at each temperature.
At a BKT transition, the measured stiffness approaches its long-distance value with logarithmic size corrections.
The stiffness curves [Fig.~\ref{fig:bkt}(a),(b)] are consistent with the critical form \cite{meth_WeberMinnhagen1988,meth_Hsieh2013}
\begin{equation}
\frac{\Ufour(L,T_c)}{T_c}
=\frac{2f}{\pi}\left(1+\frac{g}{2\ln L+C}\right),
\label{eq:bkt}
\end{equation}
where \(C\) is a nonuniversal fitting constant and \(f,g\) are fixed geometric corrections, close to unity, for a square system with periodic boundaries.
We fit \(T_c\) and \(C\) with the predicted universal jump imposed as a constraint.
Varying the minimum fitted size and the logarithmic correction gives \(T_c/t\simeq1.58\times10^{-3}\).

An independent check comes from the second-moment composite correlation length \(\xi_4\), which measures the spatial range of \(C_{4e}\).
Below the transition it grows with \(L\); at \(T/t=1.65\times10^{-3}\), it saturates near 51 lattice spacings between \(L=256\) and 512 [Fig.~\ref{fig:bkt}(c)].
The composite moment also approaches the short-range scaling established above: \(\avg{|m_4(2L)|^2}/\avg{|m_4(L)|^2}\simeq1/4\), as expected for \(\avg{|m_4|^2}\propto L^{-2}\).
Together, logarithmic stiffness scaling and a finite high-temperature correlation length support a half-vortex BKT transition from the microscopically matched \(4e\) phase to the normal network.

\textit{Half-flux response.---}
We next test whether the independent charge-\(4e\) phase exhibits the equilibrium flux response expected from its coherent charge \cite{four_Agterberg2008,four_Radzihovsky2009,four_Ge2024}.
Flux \(\Phi\) through one periodic cycle imposes an island-phase twist \(\alpha=2\pi\Phi/\Phi_0\), where \(\Phi_0=h/(2e)\) is the ordinary flux quantum.
Since \(\phi=2\theta\), the composite field experiences the doubled twist \(2\alpha\).
Adding \(\Phi_0/2\) therefore changes this twist by \(2\pi\), predicting equivalent free-energy minima separated by \(h/4e\).

\begingroup
\clubpenalty=10000
At fixed \((M,T)\), we impose the twist on the bonds crossing one periodic boundary.
Every harmonic there becomes \(-J_n(M,T)\cos[n(\theta_i-\theta_j-\alpha)]\).
The retained odd harmonics change sign under \(\alpha\to\alpha+\pi\); the full harmonic Hamiltonian remains exactly \(2\pi\)-periodic.
For the twisted-network partition function \(Z(\alpha)\), \(F(\alpha)=-T\ln Z(\alpha)\) gives \(\partial_\alpha F=\avg{\partial_\alpha H}_\alpha\), proportional to the equilibrium circulating current.
The twist free-energy cost \(\Delta F(\alpha)=F(\alpha)-F(0)\) consequently follows from
\begin{equation}
\Delta F(\alpha)=\int_0^\alpha\dd a\,
\avg{\frac{\partial H(a)}{\partial a}}_a .
\label{eq:fluxresponse}
\end{equation}
Here \(H(a)\) is the full network Hamiltonian at twist \(a\), and the average uses its thermal distribution.
\par\endgroup

\begin{figure}[tb]
\includegraphics[width=\columnwidth]{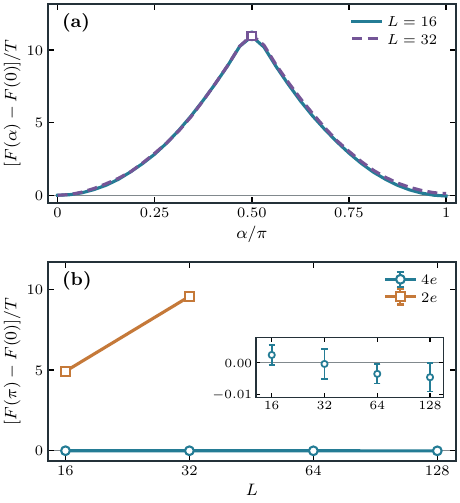}
\caption{\label{fig:flux}
Half-flux response at \(T/t=10^{-3}\).
(a) Physical twist free energies at \(M=M_*\) for \(L=16,32\), obtained by thermodynamic integration.
The points at \(\alpha=\pi/2\) use a refined integration grid.
(b) Independent \(\pi\)-twist costs: the cost in the charge-\(4e\) phase remains compatible with zero through \(L=128\) (inset), while the ordinary control at \(M/t=4.21\times10^{-2}\) has a finite cost at \(L=16,32\).
}
\end{figure}

At \(M=M_*\), \(T/t=10^{-3}\), the \(L=16,32\) profiles return close to \(F(0)\) at \(\alpha=\pi\).
Both curves peak approximately \(11T\) above \(F(0)\) near \(\alpha=\pi/2\) [Fig.~\ref{fig:flux}(a)].
This finite intermediate cost shows that the array remains sensitive to flux between the recovered minima.
Independent endpoint calculations of \(\Delta F_\pi\equiv F(\pi)-F(0)\) remain consistent with zero through \(L=128\) [Fig.~\ref{fig:flux}(b)].

\begingroup
\widowpenalty=10000
At the same temperature, a uniform charge-\(2e\) control at \(M/t=4.21\times10^{-2}\) has \(\Delta F_\pi=4.92(2)T\) and \(9.55(5)T\) for \(L=16,32\).
Changing \(M\) therefore changes both the coherent charge sector and its half-flux response.
The recovered half-flux minimum, finite intermediate cost, and ordinary-phase contrast support an emergent \(h/4e\) equilibrium response of the independent charge-\(4e\) phase.
\par\endgroup

\textit{Robustness and implications.---}
The independent charge-\(4e\) phase persists under the tested changes in junction geometry and weak microscopic disorder.
The reference microscopic junction is periodic transverse to transport; we replace this boundary by open edges.
This changes the electronic channels, so we recalculate the full harmonic spectrum and retune \(M\) to the new \(J_1=0\) point.
At \(T/t=10^{-3}\), the resulting array retains short-range ordinary correlations and algebraic composite correlations through \(L=256\).
The composite exponents inferred from size scaling and stiffness both remain near \(0.071\).

Disorder can modify the harmonics of altermagnetic junctions \cite{am_li2026disorder}.
We sample scalar on-site potentials uniformly from \([-w,w]\) inside the AM region and subtract their mean in each sample; \(w\) is the initial disorder half-width.
Each network bond receives the full harmonic spectrum of a disordered microscopic junction, allowing its couplings to vary from bond to bond.
For \(w/t=0.003,0.010\) at the same temperature, both ordinary moments retain \(L^{-2}\) scaling, while the composite moment gives \(\eta_4\simeq0.067\) through \(L=256\).
The separation of ordinary and composite coherence therefore survives the tested boundary change and weak junction inhomogeneity.

A physical array must also lie in the regime described by our classical phase model.
In an on-site capacitance model, \(E_C=(2e)^2/(2C_\Sigma)\), where \(C_\Sigma\) is the total island capacitance and \(E_C\) is the charging energy.
Harmonic phase modes then have energies \(\hbar\omega_\lambda=\sqrt{2E_Ck_\lambda}\), where \(k_\lambda\) is the curvature of the Josephson energy for mode \(\lambda\).
Requiring \(\hbar\omega_\lambda\ll T\) for the relevant modes makes their fluctuations thermal.
The magnetic screening length must also exceed the probed distances, so vortex interactions remain logarithmic over those scales \cite{meth_Pearl1964,meth_Minnhagen1987}.
These conditions specify when the classical, unscreened network description applies.

Together, the results establish altermagnetic junctions as a microscopic control of the coherent charge sector in a classical superconducting array.
The full temperature-dependent junction spectrum stabilizes an independent charge-\(4e\) phase between two ordinary superconducting states, with a thermal transition and an emergent \(h/4e\) flux response.

\begin{acknowledgments}
\textit{Acknowledgments.---}
We thank Fei Teng for his helpful discussions. This work is supported by the National Key Research and Development Program of China (Grant No. 2022YFA1404204), the National Natural Science Foundation of China (Grant No. 12274086), and the Quantum Science and Technology-National Science and Technology Major Project (Grant No. 2024ZD0300104).
\end{acknowledgments}

\nocite{am_brekke2023,am_cheng2024orientation,am_giil2024memory,am_maeland2024,am_palle2026vestigial,am_sun2023andreev,am_zhao2025orientation,am_zou2026intertwined,meth_AmbegaokarBaratoff1963,meth_Binder1981,meth_FlyvbjergPetersen1989,meth_Gao2022,meth_Josephson1962,meth_Korshunov1985,meth_Metropolis1953,meth_Shi2011,meth_SwendsenWang1987,meth_Wolff1989}
\clearpage
\bibliography{references/all_refs}
\end{document}


\title{Supplemental Material for Altermagnetic Control of Charge-\texorpdfstring{\(4e\)}{4e} Superconductivity}
\author{Yan-Guang Yue}
\affiliation{Department of Physics and State Key Laboratory of Surface Physics, Fudan University, Shanghai 200433, China}
\author{Jie Lou}
\affiliation{Department of Physics and State Key Laboratory of Surface Physics, Fudan University, Shanghai 200433, China}
\author{Yan Chen}
\email{yanchen99@fudan.edu.cn}
\affiliation{Department of Physics and State Key Laboratory of Surface Physics, Fudan University, Shanghai 200433, China}
\affiliation{Shanghai Branch, Hefei National Laboratory, Shanghai 201315, China}
\maketitle

\section{Microscopic junction and the complete harmonic spectrum}
\subsection{Hamiltonian and conventions}
Each array bond represents one independent superconductor--altermagnet--superconductor (S/AM/S) junction.
Here S and AM denote the superconducting electrodes and altermagnetic region, respectively.
This approximation requires negligible electronic propagation between distinct junction contacts on the same island, as when their separation exceeds the relevant subgap propagation length.
The island must also have sufficient internal stiffness to maintain a common condensate phase and nearly fixed pairing amplitude.
We use the electronic energy scale \(t\) and set the lattice spacing and \(k_B\) to unity.
The junction coordinate \(x\) runs from 1 to \(L_{\rm AM}\); \(y\) is transverse.
For the reference calculation, the length is \(L_{\rm AM}=20\), the transverse width is \(W=128\), and transverse boundary conditions are periodic.
The two rigid superconductors have dispersion
\(\xi_S(\bm k)=t(\cos k_x+\cos k_y)-\mu_S\), gap magnitude \(\Delta=0.01t\), and phases 0 and \(\delta\).
Both chemical potentials are \(0.2t\), and the spin-independent interface hopping is \(t_c=0.5t\).
The AM electron dispersions are
\begin{equation}
\xi_{\uparrow,\downarrow}(\bm k)=
(t\pm M)\cos k_x+(t\mp M)\cos k_y-\mu_{\rm AM}.
\end{equation}
Thus \(t/2\), rather than \(t\), is the real-space nearest-neighbor hopping in a nonmagnetic region.
This convention is necessary when translating the quoted interface hopping into a different tight-binding convention.
No pairing amplitude is imposed inside the AM.
The microscopic calculation describes a proximity junction, not an intrinsic superconducting instability of the AM.
The clean array assigns the same bond spectrum to both directions.
For the collinear model, rotating the transport direction by \(90^\circ\) interchanges the two spin dispersions; singlet electrodes make the junction energy invariant under this interchange.

For each \(k_y\), we work in the Nambu basis \((c_{\uparrow},c^\dagger_{\downarrow})\).
The AM on-site matrix and hopping matrix are
\begin{equation}
h_0(k_y)=
\begin{pmatrix}
(t-M)\cos k_y-\mu_{\rm AM}&0\\
0&-(t+M)\cos k_y+\mu_{\rm AM}
\end{pmatrix},
\qquad
V=\begin{pmatrix}(t+M)/2&0\\0&-(t-M)/2\end{pmatrix}.
\label{eq:smh}
\end{equation}
The interface hopping in Nambu space is \(t_c\tau_z\).
All momenta in the transverse Brillouin zone are included.
The real-part representation below already accounts for the second spin block and negative Matsubara frequencies.

\subsection{Lead self-energies and Schur recursion}
The exact surface Green function of a semi-infinite electrode at phase zero can be represented as
\begin{equation}
g_S(k_y,i\omega)=\frac{2}{\pi}\int_0^\pi\dd k_x\,\sin^2 k_x
\left[i\omega\tau_0-
\{t\cos k_x+t\cos k_y-\mu_S\}\tau_z-\Delta\tau_x\right]^{-1}.
\label{eq:surface}
\end{equation}
The integral can be evaluated analytically using the semi-infinite-chain resolvent.
At phase \(\theta\),
\(g_S^\theta=U_\theta g_SU_\theta^\dagger\), with \(U_\theta=\exp(i\theta\tau_z/2)\).
The boundary self-energy is
\(\Sigma_\theta=t_c^2\tau_zg_S^\theta\tau_z\).
This retains the electrode band structure and avoids a wide-band approximation.

For each \((k_y,\omega)\), initialize
\(A_1=i\omega\tau_0-h_0-\Sigma_0\).
Eliminate successive AM sites by
\begin{equation}
A_x=i\omega\tau_0-h_0-V^\dagger A_{x-1}^{-1}V,\qquad x=2,\ldots,L_{\rm AM}.
\label{eq:schur}
\end{equation}
The right electrode is added only at the last site.
All previously eliminated determinants are independent of \(\delta\) and cancel in the ratio.
For this spin-conserving junction the final determinant takes the form
\begin{equation}
D(\delta)=\det[A_{L_{\rm AM}}-\Sigma_\delta]
=P+Q\e^{i\delta}+Q\e^{-i\delta}.
\label{eq:det}
\end{equation}
Consequently,
\begin{equation}
E_J(\delta)-E_J(0)
=-2T\sum_{\omega_m>0,k_y}{\rm Re}
\ln\!\left[\frac{P+2Q\cos\delta}{P+2Q}\right],
\qquad \omega_m=(2m+1)\pi T .
\label{eq:free}
\end{equation}
The numerical implementation uses a \(\log(1+x)\) form near \(\delta=0\).
Equation~\eqref{eq:free} is the total phase-dependent junction free energy.
If the transverse sum is first divided by \(W\), the extracted coefficients must be multiplied by \(W\) before they become an array bond.
The resulting \(H_{\rm net}(T)\) is an effective classical Hamiltonian at that temperature.
Its thermal sampling and twist derivatives are well defined at fixed \(T\).
A temperature derivative of its free energy would additionally require the derivatives of \(J_n(T)\); the present work does not infer a microscopic heat capacity from \(\avg{H_{\rm net}}\).

\subsection{Analytic harmonic extraction and error control}
Let \(q=2Q/P\) and choose the root
\begin{equation}
z=-\frac{q}{1+\sqrt{1-q^2}},\qquad |z|<1.
\end{equation}
Then \(D(\delta)=A(1-z\e^{i\delta})(1-z\e^{-i\delta})\), for a phase-independent \(A\).
Expansion of both logarithms in Eq.~\eqref{eq:free} gives
\begin{equation}
J_n=-\frac{4T}{n}\sum_{\omega_m>0,k_y}{\rm Re}\,z^n,
\qquad
E_J(\delta)=-\sum_{n\ge1}J_n\cos(n\delta)+{\rm const}.
\label{eq:Jn}
\end{equation}
Direct Fourier extraction gives the same coefficients:
\begin{equation}
J_n=-\frac{1}{\pi}\int_0^{2\pi}\dd\delta\,E_J(\delta)\cos(n\delta).
\end{equation}
The current is \(I(\delta)=(2e/\hbar)\sum_n nJ_n\sin(n\delta)\) \cite{meth_Josephson1962,meth_AmbegaokarBaratoff1963}.
The factor \(n\) distinguishes current harmonics from the corresponding energy harmonics.

For a finite Matsubara sum, an absolute bound on all omitted harmonics is
\begin{equation}
\sum_{n>N_h}|J_n|
\le 4T\sum_{\omega_m>0,k_y}
\frac{|z|^{N_h+1}}{(N_h+1)(1-|z|)}.
\label{eq:tail}
\end{equation}
The microscopic expansion has no finite highest harmonic; \(N_h\) is a numerical cutoff.
All production calculations retain \(N_h=64\), providing a conservative margin for harmonic convergence.
At the reference point \(T/t=0.001\), \(M/t=0.0416847857882\), the maximum \(|z|\) is 0.47488 and the bound divided by \(T\) is \(3.65\times10^{-21}\).
The first six energy harmonics are
\begin{equation}
\frac{(J_1,J_2,J_3,J_4,J_5,J_6)}{T}
=(-5.78\times10^{-14},\,2.99454853,\,-0.13829035,\,-0.13704629,\,0.02369132,\,0.00865707).
\end{equation}
At the 13 Berezinskii--Kosterlitz--Thouless (BKT) temperatures the maximum corresponding bound is \(2.24\times10^{-31}\).
Writing \(B_{N_h}=\sum_{n>N_h}|J_n|\), the omitted bond potential obeys \(|E_J(\delta)-E_J^{(N_h)}(\delta)|\le B_{N_h}\) uniformly in \(\delta\), with the same additive constant.
For a periodic square network with \(N_s\) sites and \(2N_s\) bonds, this implies
\begin{equation}
|H-H^{(N_h)}|\le 2N_s B_{N_h},\qquad
\left|\frac{F-F^{(N_h)}}{N_s}\right|\le 2B_{N_h}.
\label{eq:truncationfree}
\end{equation}
The free-energy inequality follows by bounding the Boltzmann weights at fixed \(M,T\).
It controls the free-energy density; it does not by itself bound the error of every critical observable.
These are bounds on harmonic truncation at the chosen Matsubara cutoff; convergence of the frequency sum is a separate numerical check.
Doubling the number of Matsubara frequencies leaves the retained normalized coefficients unchanged within numerical precision.
Independent recalculation of the 13 microscopic spectra reproduces \(J_n/T\) to \(6.1\times10^{-15}\) or better.
The reference calculation uses 768 positive frequencies and a 256-point phase grid for the Fourier cross-check.

The first-harmonic root is not the exact \(0\)--\(\pi\) equal-depth point:
\begin{equation}
E_J(\pi)-E_J(0)=2\sum_{n\ {\rm odd}}J_n .
\end{equation}
At the reference root this difference is \(-0.23490T\).
The equal-depth point occurs instead near \(M/t=0.04176614\).
This distinction is relevant to a general junction potential \cite{meth_Korshunov2010} and is retained throughout the full-harmonic network calculation.

\section{Why ordinary correlations are short ranged}
Choose \(\phi_i\in[0,2\pi)\) and \(s_i=\pm1\) so that \(e^{i\theta_i}=s_i e^{i\phi_i/2}\).
For fixed director angles the network Hamiltonian becomes
\begin{equation}
H(\phi,s)=H_{\rm even}(\phi)-\sum_{\langle ij\rangle}K_{ij}(\phi)s_is_j,\qquad
K_{ij}(\phi)=\sum_{n\ {\rm odd}}J_n
\cos\!\left[\frac{n(\phi_i-\phi_j)}{2}\right].
\label{eq:ising}
\end{equation}
Different choices of the half-angle branch only relabel the signs.
There is no conditional Ising field.
Its single-site conditional distribution has
\begin{equation}
p(s_i=+1\mid s_{\ne i},\phi)
=\frac{1+\tanh[\beta\sum_jK_{ij}s_j]}{2}.
\end{equation}
Changing one neighboring sign changes this probability by at most
\(c_{ij}\le\tanh(\beta|K_{ij}|)\).
Since \(|K_{ij}|\le J_{\rm odd}\), the row sum of the influence matrix is at most
\(\mathcal D=4\tanh(J_{\rm odd}/T)\).
The Dobrushin condition \(\mathcal D<1\) gives a unique conditional sign Gibbs state and exponentially decaying sign correlations uniformly in \(\phi\) \cite{meth_Dobrushin1968}.
For example, summing paths of length at least the lattice distance \(d(i,j)\) bounds the correlation by a constant times \(\mathcal D^{d(i,j)}/(1-\mathcal D)\).
The prefactor is immaterial to the exponential decay.

The full ordinary correlation satisfies
\begin{equation}
\left|\avg{\e^{i(\theta_i-\theta_j)}}\right|
=\left|\avg{\e^{i(\phi_i-\phi_j)/2}\avg{s_is_j}_{s\mid\phi}}_{\phi}\right|
\le \avg{\left|\avg{s_is_j}_{s\mid\phi}\right|}_{\phi}.
\end{equation}
It therefore inherits the uniform exponential bound.
The argument includes both uniform and staggered ordinary channels and makes no assumption that the director sector is disordered.
At \(M=M_*\), \(T/t=0.001\), \(\mathcal D=0.65424378\); scanning \(M\) at this temperature gives the sufficient interval \(0.0416197\lesssim M/t\lesssim0.0417287\).
Its boundaries are not thermodynamic \(2e\)--\(4e\) phase boundaries.
The proof establishes only ordinary short-range correlations in the specified pairwise classical model; algebraic director correlations require the separate numerical evidence.

\section{Observables and statistical analysis}
\subsection{Correlations, structure factors, and stiffness}
For \(p=1,2\), define
\begin{equation}
C_p(\bm r)=L^{-2}\sum_i\avg{\cos[p(\theta_{i+\bm r}-\theta_i)]},\qquad
S_p(\bm k)=L^{-2}\avg{\left|\sum_i\e^{ip\theta_i+i\bm k\cdot\bm r_i}\right|^2}.
\end{equation}
The physical charges are \(2e\) for \(p=1\) and \(4e\) for \(p=2\).
For clarity the main text uses \(C_{2e}=C_1\), \(C_{4e}=C_2\).
The moments satisfy \(S_p(0)=L^2\avg{|m_p|^2}\).
The staggered ordinary channel is \(S_1(\pi,\pi)\).
An algebraic correlation \(r^{-\eta}\), integrated over a two-dimensional sample, gives \(\avg{|m|^2}\sim L^{-\eta}\).
A summable correlation gives a finite \(S(0)\) and hence \(\avg{|m|^2}\sim L^{-2}\).

Using the average over the two smallest nonzero wave vectors, the second-moment correlation length is
\begin{equation}
\xi_p=\frac{1}{2\sin(\pi/L)}
\left[\frac{S_p(0)}{\overline S_p(k_{\min})}-1\right]^{1/2},
\qquad
\overline S_p(k_{\min})=\frac{S_p(2\pi/L,0)+S_p(0,2\pi/L)}{2}.
\label{eq:xi}
\end{equation}
The associated ratio \(R_p=1-\overline S_p(k_{\min})/S_p(0)\) is useful for phase-map screening.
For the staggered channel the wave vectors are shifted by \((\pi,\pi)\).
Effective exponents between sizes \(L\) and \(2L\) are
\begin{equation}
\eta_{\rm eff}(L,2L)=
-\frac{\ln[\avg{|m(2L)|^2}/\avg{|m(L)|^2}]}{\ln2}.
\end{equation}
These are finite-size diagnostics, not automatically critical exponents.
In an ordinary spin-wave phase, \(C_2\sim C_1^4\) at long wavelengths; the relation \(\eta_4\simeq4\eta_2\) helps identify inherited composite correlations.

The helicity modulus follows by replacing each \(x\)-bond angle with \(\delta_{ij}-a_x\) and differentiating \(F\) twice at \(a_x=0\).
At fixed \(M,T\), the couplings are fixed during this derivative.
Let \(N=L^2\) and \(j_n=J_n/T\).
Then
\begin{equation}
\frac{\Upsilon_\theta^x}{T}=
\frac{1}{N}
\left[
\avg{\sum_{\langle ij\rangle_x,n}n^2j_n\cos(n\delta_{ij})}
-{\rm Var}\!\left(\sum_{\langle ij\rangle_x,n}nj_n\sin(n\delta_{ij})\right)
\right].
\end{equation}
The director free energy \(\frac12\Upsilon_4\int\dd^2r\,(\nabla\phi)^2\), with \(\phi=2\theta\), implies \(\Upsilon_4=\Upsilon_\theta/4\).
The charge-\(4e\) BKT jump is therefore expressed in terms of \(\Upsilon_4\).

\subsection{Monte Carlo updates and sampling checks}
Every update evaluates the full dimensionless potential \(V(\delta)=-\sum_nj_n\cos(n\delta)\).
Local continuous proposals use the Metropolis rule \cite{meth_Metropolis1953}, and single-site \(\pi\) flips use their exact two-state conditional probability.
For an untwisted bond, a common reflection \(\theta\mapsto2\psi-\theta\) or a common shift \(\theta\mapsto\theta+\pi\) leaves its energy unchanged.
If transforming only one end raises the dimensionless bond energy by \(\Delta_{ij}\), the neighbor is recruited with probability \(1-\exp[-\max(0,\Delta_{ij})]\).
The forward/reverse probabilities of the unoccupied boundary bonds provide detailed balance for this embedded cluster update \cite{meth_SwendsenWang1987,meth_Wolff1989}.
For a general physical twist, we use the seam acceptance correction described under \hyperref[sec:physicalflux]{Physical twist}.

One sweep consists of \(L^2\) local Metropolis proposals, each selecting a site uniformly with replacement, followed by a full pass of single-site \(\pi\) heat-bath updates.
It also includes one \(\pi\)-shift cluster and one reflection cluster, each seeded at a random site; the reflection axis \(\psi\) is uniformly random.
Every twentieth sweep adds a global proposal \(\theta_{x,y}\mapsto\theta_{x,y}\pm\pi x/L\) or \(\theta_{x,y}\pm\pi y/L\), accepted using the full energy change.
The local proposal width adapts during thermalization and remains fixed during measurement.
The four starts are uniformly random angles, an ordered configuration, random \(0/\pi\) signs, and a director-winding configuration with random signs.

The BKT simulations comprise 208 chains.
For \(L=64,128\), each chain has 40,000 thermalization sweeps and 120,000 measurement sweeps; \(L=256,512\) use 60,000 and 160,000, respectively.
Saved records are separated by 40 sweeps.
All chains use 50-record blocks, longer than eight times the largest estimated integrated autocorrelation time \(\tau_{\rm int}\), measured in saved records.
Single-chain errors use the block jackknife; pooled observables use 600 bootstrap resamples of blocks within each chain, and BKT fits use 300 draws.
Blocking accounts for correlated measurements \cite{meth_FlyvbjergPetersen1989}.

We compare the two halves of each measurement history and all pairs of independent starts, requiring agreement within 4.5 combined standard errors.
For each monitored BKT observable, we also require at least 100 effective samples, estimated as \(N_{\rm rec}/(2\tau_{\rm int})\), and blocks longer than \(8\tau_{\rm int}\).
Here \(N_{\rm rec}\) is the number of saved measurement records in the chain.
All BKT chains and size groups satisfy these criteria.
Reblocking with 100 and 200 records and varying analysis choices gives 24 fit variants with central temperatures between \(0.00157569t\) and \(0.00157670t\).

Global winding proposals have no accepted moves for \(L\ge128\).
The recorded director winding along selected lattice rows nevertheless changes in every chain, showing that these observables are not frozen.
In configurations containing vortices, row winding depends on the chosen path and does not define a unique global topological sector.
These changes and agreement between independent starts therefore do not establish complete sampling of global sectors.
We validate the implementation against exact small-system results and direct energy-difference calculations.

\clearpage
\section{Phase-map construction and representative results}
The main map uses the physical coordinates \(M/t,T/t\).
The broad scan contains 88 points and 1120 chains, followed by targeted checks of ambiguous points and the separate large-size temperature cut.
The broad scan uses 100-record blocks and checks split-half and independent-start agreement within 4.5 standard errors, together with \(\tau_{\rm int}\le10\) records.
Points failing these checks are marked as sampling inconclusive.
Among the remaining points, phase labels require consistent finite-size observables; ambiguous phase assignments are marked as unresolved.
For the colored presentation, ordinary--composite interfaces are placed between supported ordinary and independent-composite points at the same or neighboring temperature.
The upper interface uses the supported normal/composite separations and the updated fixed-\(M\) BKT estimate.
Piecewise interpolation connects these guides.
The colored boundaries have the resolution of the underlying sampling intervals and do not represent a global critical-line fit.

\begin{figure}[ht]
\includegraphics[width=0.76\textwidth]{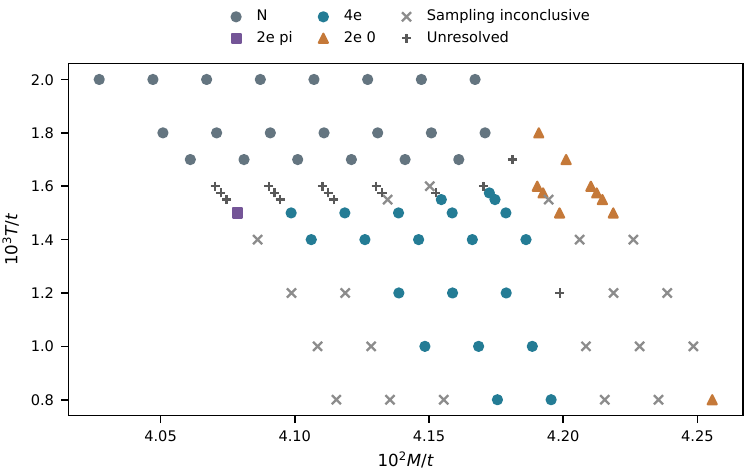}
\caption{Broad-scan sampling behind the phase-map presentation.
Symbols show finite-size phase assignments.
Gray crosses indicate inconclusive sampling; gray plus signs denote unresolved finite-size phase assignments despite satisfying the sampling criteria.
The phase boundaries use the refined ordinary brackets and large-size BKT estimate rather than the coarse-scan assignments.}
\label{fig:sampled}
\end{figure}

At \(T/t=0.001\), the left ordinary/composite separation is bracketed by \(M/t=0.04144\) and \(0.04145\), and the right separation by \(0.04204\) and \(0.04205\).
Deep in the ordinary wings the uniform or staggered exponent is approximately 0.017, while the composite exponent is approximately 0.068.
Near these boundaries, the sign-sector description suggests a relationship to criticality in generalized planar-rotor (XY) models \cite{meth_Korshunov1985,meth_Shi2011}.
These brackets locate the ordinary--composite boundaries; our finite-size crossings and Binder ratios \cite{meth_Binder1981} do not resolve their universality classes.

\begin{table}[h]
\caption{Representative large-size data at fixed \(M/t=0.0416847858\).
All stiffnesses are for the director \(\phi=2\theta\).
At \(T/t=0.00155\), the effective exponents from \(L=256,512\) are \(\eta_4=0.20925(192)\), \(\eta_{2,0}=1.995(19)\), and \(\eta_{2,\pi}=2.019(19)\).
The \(L=512\) stiffness gives \(\eta_4^\Upsilon=0.20837\). Parentheses indicate one standard error in the last quoted digits.}
\begin{ruledtabular}
\begin{tabular}{ccrrrr}
\(T/t\)&\(L\)&\(\avg{|m_4|^2}\)&\(\Upsilon_4/T\)&\(S_1(0)\)&\(S_1(\pi,\pi)\)\\
\hline
0.00155&64 &0.40478(45)&0.76899(83)&4.601(54)&0.4264(41)\\
0.00155&128&0.35003(40)&0.76583(80)&4.649(53)&0.4226(41)\\
0.00155&256&0.30250(31)&0.76428(67)&4.611(45)&0.4359(40)\\
0.00155&512&0.26166(22)&0.76381(69)&4.626(42)&0.4301(41)\\
\end{tabular}
\end{ruledtabular}
\end{table}

At \(T/t=0.001\), the finite-size scaling fit gives \(\eta_4=0.06708(28)\), consistent with the stiffness estimate \(0.06710\).

\clearpage
\section{BKT finite-size analysis and director defects}
For a unit director vortex, the smooth-phase energy is
\(E_v=\pi\Upsilon_4\ln(L/a)+E_{\rm core}\), while its positional entropy is \(2\ln(L/a)\).
The balance motivates the renormalized jump \(\Upsilon_4^R/T=2/\pi\) \cite{meth_KosterlitzThouless1973,meth_NelsonKosterlitz1977}.
At criticality the running vortex fugacity gives logarithmic size corrections \cite{meth_Kosterlitz1974,meth_WeberMinnhagen1988,meth_Hsieh2013}.
We linearly interpolate each size's measured stiffness and its uncertainty between adjacent sampled temperatures and fit
\begin{equation}
\frac{\Upsilon_4(L,T_c)}{T_c}
=\frac{2f}{\pi}\left[1+\frac{g}{2\ln L+C+\mathcal L(L,C)}\right],
\end{equation}
with either \(\mathcal L=0\), as displayed in the Letter, or \(\mathcal L=\ln(\ln L+C/2)\).
The square-torus factors are \(f=0.99982471\) and \(g=1.00202783\).
The fitted parameters are \(T_c\) and \(C\); the jump and torus factors are theoretical inputs.
We restrict \(T_c/t\) to the sampled interval \([0.00150,0.00165]\) and \(C\) to \([2(1.1-\ln L_{\min}),80]\), keeping the logarithmic denominators positive.
Fits start from \(C=0,4,12\) and three temperatures within this interval.
All four central fits and their 300 bootstrap draws lie inside these bounds.
Therefore a successful fit is a test of BKT finite-size behavior, not an independent extraction of the jump's universal numerical value.

\begin{table}[h]
\caption{BKT estimates from the full microscopic temperature sweep.
Their spread measures part of the systematic uncertainty. Including the bootstrap uncertainty gives the working interval \(0.001575\le T_c/t\le0.001578\).}
\begin{ruledtabular}
\begin{tabular}{ccrrr}
\(L_{\min}\)&logarithmic form&\(10^3T_c/t\)&\(C\)&\(\chi^2/{\rm dof}\)\\
\hline
64 &leading&1.5756983&2.35169&1.31018/2\\
64 &next logarithm&1.5760793&1.00000&0.89550/2\\
128&leading&1.5762487&3.05393&0.00748/1\\
128&next logarithm&1.5765205&1.55048&0.00637/1\\
\end{tabular}
\end{ruledtabular}
\label{tab:bkt}
\end{table}
Across these fits, the smallest lower and largest upper bootstrap endpoints are
\(0.0015748986t\) and \(0.0015775516t\).
The small number of degrees of freedom at \(L_{\min}=128\) limits how discriminating its reduced residual can be.
Agreement across both size cuts and both forms is more informative than its particularly small \(\chi^2\).
The square-torus correction shifts the fitted temperatures only slightly; removing it in the leading fit changes \(T_c/t\) from 0.0015756983 to 0.0015756775.

As a distinct high-temperature check, at \(T/t=0.00165\), \(\xi_4=51.78(41)\) for \(L=256\) and \(51.15(86)\) for \(L=512\).
The moment ratio \(0.25394\) gives \(\eta_{\rm eff}=1.9775(157)\), consistent with short-range composite correlations.
At the largest system size, the residual stiffness estimate is small, with \(\Upsilon_4/T\simeq0.011\).

\begin{figure}[t]
\includegraphics[width=0.91\textwidth]{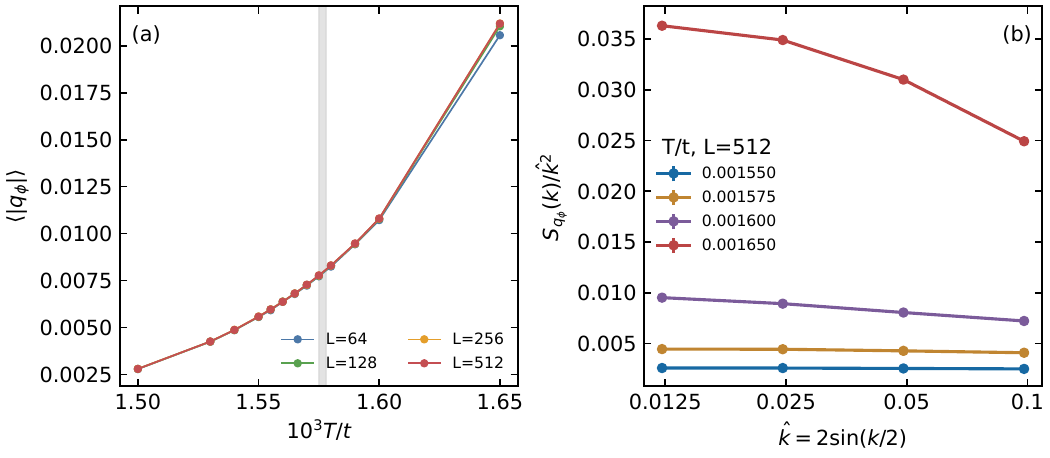}
\caption{\label{fig:defects}Director-defect diagnostics from the large-size temperature sweep.
Unit charge refers to winding of \(\phi=2\theta\), corresponding to a half vortex of the original phase.
Finite core density is compatible with bound pairs below the transition and cannot by itself diagnose unbinding.}
\end{figure}

For each plaquette define the director charge
\begin{equation}
q_\phi=\frac{1}{2\pi}\sum_{\partial p}
{\rm Arg}\!\left[\e^{2i(\theta_j-\theta_i)}\right].
\end{equation}
The cores with \(q_\phi=\pm1\) are half vortices of the original phase modulo its sign branch.
In an ordinary phase such defects may be connected by sign-domain walls; identifying cores alone does not show that they are unbound.
The measured density rises from approximately 0.0056 at \(T/t=0.00155\) to 0.0212 at \(T/t=0.00165\) for \(L=512\), with a value near 0.0078 around the transition.
The charge structure factor averages the two lattice directions:
\begin{equation}
S_{q_\phi}(k)=\frac{1}{2L^2}\left\langle
\left|\sum_p q_\phi(p)e^{ikx_p}\right|^2+
\left|\sum_p q_\phi(p)e^{iky_p}\right|^2
\right\rangle .
\end{equation}
Here \(p\) labels plaquettes, and \((x_p,y_p)\) is the coordinate of their lower-left corner.
We evaluate \(k=2\pi m/L\) for \(m=1,2,4,8\); Fig.~\ref{fig:defects}(b) shows \(S_{q_\phi}(k)/\hat k^2\), where \(\hat k=2\sin(k/2)\).
Density and charge-structure-factor trends support the unbinding interpretation based on the combined BKT and correlation evidence.

\section{Physical flux and auxiliary seam integration}\label{sec:physicalflux}
\subsection{Physical twist}
On a periodic array, select all positively oriented \(x\) bonds crossing one seam and replace their phase difference by \(\delta_{ij}-\alpha\).
The gauge-invariant twist is \(\alpha=2\pi\Phi/\Phi_0\), with \(\Phi_0=h/(2e)\).
Only the seam is shifted; all harmonics transform as
\(\cos[n(\delta_{ij}-\alpha)]\).
At general \(\alpha\), a common reflection changes the energy of a seam bond even when both endpoints are reflected.
We therefore build reflection clusters using only the bulk bonds and accept the proposed reflection with probability
\begin{equation}
P_{\rm acc}=\min\{1,\exp[-\Delta H_{\rm seam}/T]\}.
\end{equation}
The energy difference includes all seam bonds, including those with both endpoints inside the cluster.
The bulk proposal ratio supplies \(e^{-\Delta H_{\rm bulk}/T}\), and the acceptance ratio supplies \(e^{-\Delta H_{\rm seam}/T}\), giving detailed balance for the full Hamiltonian.
At \(\alpha=0,\pi\), the seam potential is even and the reflection cluster can include all bonds directly.
The \(\pi\)-shift cluster remains valid at every \(\alpha\), since shifting both endpoints leaves their phase difference unchanged.
For the \(L=16,32\) profiles, each angle uses four starts with 8,000 thermalization and 32,000 measurement sweeps, saving a record every 20 sweeps.

The free-energy difference obeys
\begin{equation}
\frac{F(\alpha)-F(0)}{T}
=\int_0^\alpha\dd a\,
\avg{\frac{1}{T}\frac{\partial H(a)}{\partial a}}_a .
\label{eq:fluxint}
\end{equation}
An individual finite full-odd-harmonic array is exactly \(2\pi\) periodic.
The emergence of a \(\pi\) period is assessed from its equilibrium free-energy profile and the residual cost at \(\pi\), rather than imposed as an exact identity.
\mbox{The central \(L=16,32\)} profiles have a maximum near \(\pi/2\).
Refinement of the integration grid near this maximum gives \(\Delta F(\pi/2)/T=10.985(92)\) and \(10.972(153)\) for \(L=16,32\), respectively.

\subsection{A lower-noise endpoint calculation}
To measure only \(F(\pi)-F(0)\), retain the even seam potential and interpolate the odd seam potential:
\begin{equation}
H_\lambda=
H_{\rm bulk}+H_{{\rm seam,even}}+(1-2\lambda)H_{{\rm seam,odd}},
\qquad 0\le\lambda\le1.
\end{equation}
Here \(H_{{\rm seam,odd}}=-\sum_{{\rm seam},\,n\ {\rm odd}}J_n\cos(n\delta_{ij})\).
The endpoints equal the untwisted and physical \(\pi\)-twisted Hamiltonians.
At every \(\lambda\), the bond potentials remain even in the phase difference, so the reflection cluster uses all bonds without a seam acceptance correction.
The intermediate \(\lambda\) configurations are an auxiliary thermodynamic path.
They do not represent \(\alpha=\pi\lambda\).
Thus
\begin{equation}
\frac{F(\pi)-F(0)}{T}
=-2\int_0^1\dd\lambda\,\avg{H_{{\rm seam,odd}}/T}_\lambda .
\label{eq:lambda}
\end{equation}
We evaluate the integral on the grid \(\lambda=k/16\), \(k=0,\ldots,16\), with independent starts.
Small-system direct enumeration checks the endpoint integral; one test gives 1.5114013 exactly and 1.5114033 by 17-point Simpson integration.
For a purely even potential the endpoint difference vanishes identically, providing another implementation check.

For both integration paths, block lengths are \(\max(50,\lceil20\tau_{\rm int}\rceil)\) saved records; we require at least eight blocks and \(\tau_{\rm int}\le40\) records.
The split-half difference is compared with \(4.5\times2\sigma\), where \(\sigma\) is the blocked standard error of the full-chain mean.
Independent-start means must agree within 4.5 combined standard errors.

\begin{table}[h]
\caption{Endpoint twist costs at \(T/t=0.001\).
Quoted uncertainties are statistical errors from the reported endpoint analyses.
Small negative estimates in the composite state are consistent with zero.
The ordinary control uses \(M/t=0.0421\).}
\begin{ruledtabular}
\begin{tabular}{crr}
\(L\)&\(\Delta F_\pi/T\), composite \(M=M_*\)&\(\Delta F_\pi/T\), ordinary control\\
\hline
16 & \(0.00248(317)\)&\(4.9179(194)\)\\
32 & \(-0.00037(477)\)&\(9.5532(539)\)\\
64 & \(-0.00352(309)\)& ---\\
128& \(-0.00460(453)\)& ---\\
\end{tabular}
\end{ruledtabular}
\end{table}
The ordinary-phase twist calculations beyond the free-energy barrier and the endpoint calculations at larger sizes do not satisfy these sampling criteria.
We use only the reliable \(L=16,32\) endpoint results in the quantitative comparison and do not extract a full ordinary-phase twist profile or an asymptotic domain-wall tension.
This response is an equilibrium signature of the classical network and does not include a ring's self-inductance.

\section{Open edges and microscopic disorder}
Opening the junction's transverse boundary changes its channel spectrum.
The reference first-harmonic zero is then retuned to \(M/t\simeq0.04215360\) at \(T/t=0.001\).
The resulting full-harmonic network gives \(\eta_4^{\rm size}\simeq0.07138\) and
\(\eta_4^\Upsilon\simeq0.07178\) through \(L=256\), with ordinary short-range correlations.
The calculation therefore tests the survival of the mechanism under a boundary change, rather than requiring its original tuning value to remain fixed.

For each microscopic disorder sample, we draw \(u_{xy}\) independently from \(U[-w,w]\) on an \(L_{\rm AM}\times P\) supercell with \(P=32\).
We then set \(V_{xy}=u_{xy}-\overline u\), where \(\overline u=(L_{\rm AM}P)^{-1}\sum_{xy}u_{xy}\), and repeat the potential with transverse period \(P\).
Thus \(w\) is the half-width before mean subtraction, which holds the sample-averaged chemical potential fixed.
All samples use the same clean reference value \(M=M_*\) at \(T/t=0.001\), without retuning individual junctions or disorder ensembles.
The junction retains width \(W=128\): four Bloch sectors of the period-32 supercell include all 128 transverse states.
Their summed free energy gives the full-junction harmonics; coefficients evaluated per transverse site are multiplied by \(W\) before entering the network.
We generate 64 junction spectra for each of \(w/t=0.003,0.010\).
Each network bond independently selects a spectrum from this bank.
For each disorder width we simulate 12 network realizations, four sizes \(L=32,64,128,256\), and four initial conditions.
Disorder realizations, not individual measurements within a chain, are the independent samples for the disorder average.
For \(w/t=0.010\), the bank spans \(J_1/T=-0.0354\)--0.0757 and \(J_2/T=2.9626\)--2.9954.
It therefore probes weak effective bond disorder.

\begin{table}[h]
\caption{Composite exponents in the microscopic disorder controls at \(T/t=0.001\).
The brackets are 95\% bootstrap intervals for the size fit; stiffness estimates use \(L=256\).
All four sizes enter the fit.}
\begin{ruledtabular}
\begin{tabular}{crrr}
\(w/t\)&\(\eta_4^{\rm size}\)&95\% interval&\(\eta_4^\Upsilon\)\\
\hline
0&0.06808&[0.06548,0.07027]&0.06711\\
0.003&0.06726&[0.06661,0.06798]&0.06714\\
0.010&0.06730&[0.06661,0.06797]&0.06761\\
\end{tabular}
\end{ruledtabular}
\end{table}
The disorder simulations comprise 448 chains including the clean controls.
Five chains show discrepancies between the first and second halves of their measurement histories.
Removing affected realization--size groups as a sensitivity check changes the two disordered exponents to 0.06724 and 0.06733; the reported main estimates retain the full sample.
The ordinary uniform and staggered moments have exponents close to 2.
The disordered calculation retains 64 harmonics, but its omitted-harmonic tail has not been bounded by the analytic clean-junction formula.
Accordingly, the ordinary short-range conclusion for these disorder controls is numerical.
Our robustness conclusions apply to the tested weak scalar disorder and transverse boundary modification.

\section{Conditions for a classical device interpretation}
The calculation treats the superconducting phases as classical coordinates.
To determine the regime of validity of the classical approximation, we consider an on-site charging term \(E_C\sum_i n_i^2\), where
\(E_C=(2e)^2/(2C_\Sigma)\) and \([\theta_i,n_j]=i\delta_{ij}\).
Expanding the phase potential around a stable configuration and diagonalizing its Hessian gives
\begin{equation}
H_{\rm harmonic}=\sum_\lambda
\left(E_C n_\lambda^2+\frac{k_\lambda}{2}\vartheta_\lambda^2\right),
\qquad
\hbar\omega_\lambda=\sqrt{2E_Ck_\lambda}.
\end{equation}
For each nonzero mode \(k_\lambda>0\), the oscillator variance is
\begin{equation}
\avg{\vartheta_\lambda^2}
=\sqrt{\frac{E_C}{2k_\lambda}}\,
\coth\!\left(\frac{\sqrt{2E_Ck_\lambda}}{2T}\right)
\longrightarrow\frac{T}{k_\lambda}
\quad\text{for }\hbar\omega_\lambda\ll T .
\end{equation}
The uniform phase mode has \(k_\lambda=0\) and is a free global rotation, excluded from this oscillator expression.
A sufficient condition for a fully thermal harmonic regime is therefore
\(2E_Ck_{\max}\ll T^2\), or \(C_\Sigma\gg(2e)^2k_{\max}/T^2\), when energies and capacitances are expressed in consistent physical units.
The weaker inequality \(E_C\ll J_2\) suppresses strong quantum fluctuations but does not, by itself, establish this thermal limit.
For a nonlocal capacitance matrix, \(H_C=(2e)^2\sum_{ij}n_i(C^{-1})_{ij}n_j/2\).
Its mode spectrum must be evaluated for the specified geometry.
These conditions delimit the applicability of the classical model.

The unscreened BKT description also requires logarithmic vortex interactions over the observed distances \cite{meth_Pearl1964,meth_Minnhagen1987}.
For a homogeneous thin film, the Pearl scale is \(\Lambda=2\lambda_L^2/d\).
An island array requires the screening length derived from its actual inductance network.
For a finite device, a sufficient spatial window is a screening length larger than both the measured correlation length and the array size in physical units.
Finite screening changes the asymptotic vortex interaction and limits extrapolation to an infinite unscreened array.
No value in kelvin is assigned to the quoted \(T/t\) until a microscopic hopping scale and a junction geometry are specified.

\section{Scope of the contribution}
The microscopic model describes a compensated, collinear spin splitting, with spin-conserving interfaces and conventional singlet electrodes.
Andreev reflection \cite{am_sun2023andreev}, magnon-mediated pairing \cite{am_brekke2023,am_maeland2024}, and memory functionality \cite{am_giil2024memory} illustrate complementary links between altermagnetism and superconductivity.
Orientation-dependent transport has also been studied with unconventional superconducting electrodes \cite{am_cheng2024orientation,am_zhao2025orientation}.
Intertwined orders in metallic altermagnets \cite{am_zou2026intertwined} and competing vestigial orders \cite{am_palle2026vestigial} concern different order-parameter settings.
Here the composite field is \((e^{i\theta})^2=e^{2i\theta}\); the amplitude of each island's ordinary condensate remains rigid.

Paired phases of generalized-XY models \cite{meth_Korshunov1985,meth_Shi2011,meth_Gao2022} and magnetic \(0\)--\(\pi\) arrays \cite{meth_Korshunov2010} are established precedents.
Here the independent charge-\(4e\) phase emerges from collective ordering with the complete microscopic altermagnetic Josephson spectrum.
Residual odd harmonics shift the equal-depth point and constrain ordinary coherence.
The BKT, flux, and robustness calculations retain the full microscopic harmonic spectrum appropriate to each parameter set and junction configuration.

The results describe a classical, unscreened array with rigid superconducting amplitudes.
A material-specific AM tuning protocol and the quantum phase diagram remain open questions.

\clearpage
\bibliography{references/all_refs}